\documentstyle[12pt]{article}
\begin{document}
\title{Note on restoring manifest
rotational symmetry in hyperfine and fine structure in light-front QED.}
\author{ Martina Brisudov\' a  and Robert Perry \\
{\it Department of Physics}\\ {\it The Ohio State University,
 Columbus, OH 43210.} }
\date{\today}
\maketitle
\abstract{We study the part of the renormalized, cutoff QED light-front 
Hamiltonian that does not change particle number.
The Hamiltonian contains interactions that 
must be treated in second-order bound state perturbation theory to obtain 
hyperfine structure. We show that a simple unitary transformation leads 
directly to the familiar Breit-Fermi spin-spin and tensor interactions, 
which can be treated in degenerate first-order bound-state perturbation 
theory, thus simplifying analytic light-front QED calculations. To the 
order in momenta we need to consider, this transformation is 
equivalent to a Melosh rotation. We also study how the similarity 
transformation affects spin-orbit interactions.
}
\vskip .25in
\newpage

\section{Introduction}

Light-front Hamiltonian field theory is being developed as a tool for 
solving
bound state problems in QCD \cite{thelongpaper,P2}.  Cutoffs are introduced
 that can
be lowered using a similarity renormalization group \cite{similarity}, and
renormalization can be completed either using coupling coherence
 \cite{couplcoh}   or by fixing
counterterms to repair symmetries violated by the cutoffs.
In the simplest procedure the renormalized, cutoff Hamiltonians are computed perturbatively and may 
then be diagonalized non-perturbatively to obtain low-lying bound states.  
Each stage has an  approximation scheme associated with it: in the first step, 
the effective Hamiltonian is calculated to a given order in perturbation 
theory; and in the second step the effective Hamiltonian is divided into a 
dominant part, which defines the starting bound-state wave functions,
 and a perturbation, which is treated to a given order using bound-state 
perturbation theory.  Interactions that change particle number are treated 
perturbatively.  
As a consequence, different Fock states decouple and 
one is left with few-body problems in the leading order.  
It is therefore important to know to what extent  
bound states are accurately described 
by the  
effective interactions that do not  change particle 
number. A principal signature of the
truncation errors in these schemes 
is a violation of rotational invariance, which is a 
dynamical symmetry in light-front field theory. 

There are two reasons why  rotational symmetry is complicated in this approach. 
The first is that we have formulated the theory 
on the light front.  Rotations are dynamical, and light-front spinors 
depend on the choice of z-axis. In the weak-coupling 
limit, however, one expects that rotations should become 
simple because both boost and rotational symmetries are kinematic in the 
nonrelativistic limit. The second source of complexity
 is the regularization and 
renormalization
 scheme that we use \cite{thelongpaper,P2}. 
 This has profound effects in QCD \cite{US}, 
but 
in QED, to obtain the leading 
 interactions cutoffs can effectively be removed. 
Therefore, to disentangle the two problems it is useful to study QED in the 
nonrelativistic limit.

Jones et al \cite{trulko} have  studied the ground state hyperfine structure
of positronium in this approach. 
To  second order in the coupling there are new interactions 
between 
electron and positron that arise from eliminating matrix elements of the 
Hamiltonian involving high-energy photon emission and absorption. 
These new interactions 
have spin-independent  as well as spin-dependent parts. 
The spin-independent interaction, combined with the instantaneous exchange 
interaction, leads to the Coulomb interaction,
 and the leading-order 
problem  reduces to the familiar equal-time Schr{\"o}dinger equation \cite{P2}.

The light-front spin-dependent
interactions appear to be  
different from the spin-dependent interactions found in an  
equal-time formulation, even in the nonrelativistic limit.  
For example, if one calculates the hyperfine splitting in the positronium 
ground state, first-order bound-state perturbation theory gives 
incorrect results. The triplet state is not degenerate, and the energy of 
the singlet state is incorrect. This is because the effective Hamiltonian 
contains a term that does not give a contribution in  first-order
 bound-state perturbation theory, but its contribution in second-order 
bound-state 
perturbation theory is of the same order in $\alpha$ as the terms 
contributing in the first order, ${\cal O}(\alpha^4)$.
Jones et al \cite{trulko} effectively
summed the second-order bound-state perturbation 
theory analytically, and showed that it leads to the correct hyperfine 
splitting in the ground state of positronium. 
Kaluza and Pirner encountered the same problem \cite{mlaka}, and they
completed the sum   numerically.

We propose  an alternative approach. We find a simple 
unitary transformation of the Hamiltonian
that alters the problematic term 
so that it enters at  first-order in 
bound-state perturbation theory. We find the transformation order-by-order 
in powers of momenta. In order to restore the hyperfine splitting, we need 
to find the unitary transformation only to the next-to-leading order.
A  unitary transformation does not change the eigenvalues, and the 
transformation we obtain  makes 
the calculation much easier. 
It turns out that the unitary transformation to this order is an expansion
of the so called Melosh transformation \cite{melosh} to next-to-leading 
order in powers of momenta. 
The simple formalism enables us to study how the similarity transformation 
affects the spin-dependent structure of the effective Hamiltonian. This 
issue is addressed in the third section. The last section contains our 
conclusions.

\section{Spin-spin interaction in QED}
The effective Hamiltonian for a fermion and an antifermion
 generated by the 
similarity transformation \cite{similarity} using 
coupling coherence
 \cite{couplcoh} is
\begin{eqnarray}
H_{\rm eff} = H_{\rm free} + V_1 +  V_{2} +  V_{2 \ {\rm eff}} \  \  ,
\end{eqnarray}
where $H_{\rm free}$ is the kinetic energy, $V_1$ is ${\cal O}(g)$ 
emission and absorption, $V_{2} $ is
the ${\cal O}(g^2)$
instantaneous exchange interaction,  and $V_{2 \ {\rm eff}}$ includes 
the ${\cal O}(g^2)$
effective interactions. 
For simplicity, we only  consider the case where the fermion and 
antifermion  have equal mass.

Kinetic energy is diagonal in momentum space, and 
matrix elements of the interactions are nonzero 
only between states with energy difference smaller than  
${\Lambda^2\over{{\cal P}^+}}$, which is reflected by an overall cutoff 
function in the equations below.
If the cutoff is 
chosen within certain limits, 
the cutoff functions can be approximated by $1$ to leading order in 
$\alpha$ \cite{P2}.

Matrix elements of the Hamiltonian in a state containing a 
fermion and   antifermion pair are as follows. 
The kinetic energy is diagonal in 
 momentum space:
\begin{eqnarray*}
{p^{\perp 2}+m^2\over{p^+}}+{k^{\perp 2}+m^2\over{k^+}} \ \ \  .
\end{eqnarray*}
The emission and absorption of a photon 
enters at second order.

Let $p_i$, $k_i$ be the light-front three-momenta carried 
by the fermion and antifermion; 
$\sigma _i$, $\lambda_i$ are their light-front helicities; 
$u(p, \sigma )$, $v(k, \lambda )$ are their spinors; 
index $i=1,2$ refers to the initial and final states, respectively.
The instantaneous exchange interaction mixes states of different momenta:
\begin{eqnarray}
  & -g^2 \bar{u}(p_2, \sigma _2) \gamma ^{\mu} u(p_1, \sigma _1)
\bar{v}(k_1, \lambda_1) \gamma ^{\nu} v(k_2, \lambda _2) 
\nonumber\\
 & \times
 {1\over{{q^+}^2}} \eta_{\mu}\eta_{\nu} \ 
\theta \left({\Lambda^2 \over{{\cal P}^+}} - \vert (p_1^- + k_1^- )
- (p_2^- +k_2^-)\vert \right) \ \ \  ,
\end{eqnarray}
and so do the effective interactions generated by the similarity 
transformation:
\begin{eqnarray}
  & -g^2 \bar{u}(p_2, \sigma _2) \gamma ^{\mu} u(p_1, \sigma _1)
\bar{v}(k_1, \lambda_1) \gamma ^{\nu} v(k_2, \lambda _2) 
\nonumber\\
 & \times {1\over{q^+}} 
D_{\mu \nu}(q)
\left({\theta(\vert D_1\vert-{\Lambda^2 \over{{\cal P}^+}})
\theta(\vert D_1\vert -\vert D_2\vert )\over{D_1}} 
+ {\theta(\vert D_2\vert-{\Lambda^2 \over{{\cal P}^+}})
\theta(\vert D_2\vert -\vert D_1\vert )\over{D_2}}\right)  
\nonumber\\
& \times \theta \left({\Lambda^2 \over{{\cal P}^+}} - \vert (p_1^- + k_1^- )
- (p_2^- +k_2^-)\vert \right)
  .
\end{eqnarray}
\noindent where 
$D_{\mu \nu}(q) = {{q^{\perp}}^2\over{{q^+}^2}}\eta_{\mu}\eta_{\nu}
 + {1\over{q^+}}
\left(\eta_{\mu}{q^{\perp}}_{\nu} + \eta_{\nu}{q^{\perp}}_{\mu}\right)
- g^{\perp}_{\mu \nu}$ is the photon propagator in light-front gauge, 
$\eta _\mu = ( 0, \eta _+ = 1,0,0)$;
$q=p_1 - p_2$ is the exchanged momentum, with
$q^- ={ {q^{\perp}}^2\over{q^+}}$;
$D_1$, $D_2$ are energy denominators:
$D_1 = p_1^- -p_2^- -q^-$ and $D_2 = k_2^- -k_1^- - q^-$.
It is convenient to add $(2)$ and $(3)$ together, leading to:
\begin{eqnarray}
\lefteqn{
 g^2 \bar{u}(p_2, \sigma _2) \gamma ^{\mu} u(p_1, \sigma _1)
\bar{v}(k_2, \lambda_2) \gamma ^{\nu} v(k_1, \lambda _1) }
\nonumber\\
& \times & \left[ g_{\mu \nu}
\left(
{\theta(\vert D_1\vert-{\Lambda^2 \over{{\cal P}^+}})
\theta(\vert D_1\vert -\vert D_2\vert )\over{q^+ D_1}} 
+ {\theta(\vert D_2\vert-{\Lambda^2 \over{{\cal P}^+}})
\theta(\vert D_2\vert -\vert D_1\vert )\over{q^+ D_2}}\right) 
\right. \nonumber\\
& & \left.
-{\eta_{\mu} \eta_{\nu} \over{2{q^+}^2}}
\left( 1 - 
{\theta(\vert D_1\vert-{\Lambda^2 \over{{\cal P}^+}})
\theta(\vert D_1\vert -\vert D_2\vert )D_2\over{D_1}} 
-{\theta(\vert D_2\vert-{\Lambda^2 \over{{\cal P}^+}})
\theta(\vert D_2\vert -\vert D_1\vert ) D_1\over{D_2}}\right) 
\right] 
\nonumber\\
& \times &  \theta \left({\Lambda^2 \over{{\cal P}^+}} - \vert (p_1^- + k_1^- )
- (p_2^- +k_2^-)\vert \right)
  .
\end{eqnarray}
In what follows, we approximate the cutoff functions by 1, which is allowed 
to leading order for the  range of cutoffs 
$g^2 m^2 \ll \Lambda ^2 \ll g m^2$ \cite{P2}.

The ``$\eta_{\mu} \eta_{\nu}$'' term is spin independent, it vanishes on 
shell, and it is at least 
one power of momenta higher than the leading spin-independent piece of the 
``$g_{\mu \nu}$'' term. 
As we explain later, it does not affect spin-spin and tensor 
interactions, but it may influence the spin-orbit.

In what follows we use Jacobi momenta:
\begin{eqnarray*}
p_i^+  =  x_i P^+ &, & p_i^{\perp}  =  \kappa _i ^{\perp} \  \  , \nonumber
\\
k_i^+  =  y_i P^+ &, & k_i^{\perp}  =  - \kappa _i ^{\perp}  \  \  ,
\end{eqnarray*}
where $y_i=1-x_i$;
and we replace four-component spinors $u(p,\sigma)$, $v(k,\lambda)$
 with two-component spinors by 
substituting:
\begin{eqnarray}
u(p,\sigma) =\sqrt{ {2\over{p^+}} } (p^+ + \beta m 
- \vec{\alpha}^{\perp}\cdot \vec{p}_{\perp} ) \Lambda _+
\left( \begin{array}{c} \xi_{\sigma} \\ 
0
\end{array} \right),
\end{eqnarray}
and similarly $v(k, \lambda)$.
Here $\beta = \gamma ^0$, $\vec{\alpha} = \gamma^0 \vec{\gamma}$ are Dirac 
matrices, $\Lambda _+ = {1\over{4}} \gamma ^- \gamma ^+$ is a projection 
operator, and $\xi _{\sigma} $ is a two-component spinor,
\begin{eqnarray*}
\xi _{\uparrow} = \left( \begin{array}{c} 1\\ 0 \end{array}\right) &  , &  
\xi _{\downarrow} =\left( \begin{array}{c} 0\\ 1 \end{array} \right) \  \  \  .
\end{eqnarray*}
 From now on we will write the Hamiltonian as 
an operator which acts in the cross product space of these 
two-component spinors.

After the following change of variables, which defines $p_z$,
\begin{eqnarray}
x_i = { \sqrt{\vec{p_i}^2 +m^2} + p_{i z} \over{2 \sqrt{\vec{p_i}^2 +m^2} }}
,
& \  &
y_i = { \sqrt{\vec{k_i}^2 +m^2} + k_{i z} \over{2 \sqrt{\vec{k_i}^2 +m^2} }},
\end{eqnarray}
where $\vec{k_i}=-\vec{p_i}$, and the three momentum in the center 
of mass frame is then $\vec{p}\equiv (\kappa ^{\perp}, p_z)$, we take the 
nonrelativistic 
limit of the Hamiltonian. 

The energy denominators become:
\begin{eqnarray}
\lefteqn{q^+ D_1   =}\nonumber\\
 & &  -{\vec{q_{  }}} ^2 + 
\left({p_{1  {\rm z}}\over{ m \sqrt{1+{\vec{p_1}^2 \over{m^2}}}}}
-{p_{2  {\rm z}}\over{ m \sqrt{1+{\vec{p_2}^2 \over{m^2}}}}}\right) 
(\vec{p_1}^2 -\vec{p_2}^2)
+ \left[{ 2 +{\vec{p_1}^2 \over{m^2}}+{\vec{p_2}^2 \over{m^2}}
\over{   \sqrt{ 1+{\vec{p_1}^2 \over{m^2}}} 
\sqrt{1+{\vec{p_2}^2 \over{m^2}}} }}
-2\right] 
p_{1 {\rm z}} p_{2  {\rm z}} \nonumber\\
 &  & \simeq - {\vec{q_{  }}}^2 , \nonumber\\
\lefteqn{q^+ D_2   =} \nonumber\\
 & &  -\vec{q_{ }}^2 -
\left({p_{1  {\rm z}}\over{ m \sqrt{1+{\vec{p_1}^2 \over{m^2}}}}}
-{p_{2  {\rm z}}\over{ m \sqrt{1+{\vec{p_2}^2 \over{m^2}}}}}\right) 
(\vec{p_1}^2 -\vec{p_2}^2)
+ \left[{ 2 +{\vec{p_1}^2 \over{m^2}}+{\vec{p_2}^2 \over{m^2}}
\over{   \sqrt{ 1+{\vec{p_1}^2 \over{m^2}}} 
\sqrt{1+{\vec{p_2}^2 \over{m^2}}} }}
-2\right] 
p_{1  {\rm z}} p_{2  {\rm z}} \nonumber\\
 & & \simeq - \vec{q_{ }}^2  ,
\end{eqnarray}
and the interaction Hamiltonian reduces to 
\begin{eqnarray}
4 g^2 (2m)^2 \left[{1\over{- {\vec{q_{ }}}^2 }} (1+d)\right]  (v_0 + v_{spin})
+ v_{\eta _{\mu} \eta _{\nu}},
\end{eqnarray}
where $v_0$ and $v_{spin}$ come from the $g_{\mu \nu}$ term in eqn. $(4)$. 
 $v_0$ is spin-independent and $v_{spin}$ depends on spins. $d$ 
denotes corrections from energy denominators that we discuss in the next 
section, together with the spin independent $v_{\eta _{\mu} \eta _{\nu}}$ that 
arises from the $\eta_{\mu} \eta _{\nu}$ term in eqn. $(4)$. We have 
dropped an 
overall factor of $\sqrt{x_1 x_2(1-x_1)(1-x_2)}$ in the Hamiltonian 
which is absorbed by a similar 
factor in the definition of the two-body  wave function.

The corrections from energy denominators  do not influence the 
discussion of spin-dependent structure, because they enter as an overall 
factor multiplying the entire $g_{\mu \nu}$ 
term. This will become clear later.
To second order in powers of momenta,
\begin{eqnarray}
v_0 = 1  + {1\over{4m^2}}(\vec{p}_1+\vec{p}_2)^2 
+ {1\over{2m^2}}\vec{p}_1 \cdot \vec{p}_2
+{1\over{2m^2}}\vec{p}_1^{\perp} \cdot \vec{p}_2^{\perp} 
 + {3\over{4m^2}}\left((\vec{p_1})_{z}^2+(\vec{p_2})_{z}^2\right) ,
\end{eqnarray}
and
\begin{eqnarray}
v_{spin}= -{{\rm i}\over{2m}} ((\vec{k}_1-\vec{k}_2) \times \vec{\sigma}_a)_{z}
-{{\rm i}\over{2m}} ((\vec{p}_1-\vec{p}_2) \times \vec{\sigma}_b)_{z} \nonumber\\
 + {1\over{4 m^2}} 
((\vec{k}_1-\vec{k}_2) \times \vec{\sigma}_a)_{\perp}\cdot 
((\vec{p}_1-\vec{p}_2) \times \vec{\sigma}_b)_{\perp} \nonumber\\
 +3 { {\rm i}\over{4 m^2}}
(\vec{k}_2 \times \vec{k}_1) \cdot   \vec{\sigma}_a
+ 3 { {\rm i}\over{4 m^2}}
(\vec{p}_2 \times \vec{p}_1) \cdot   \vec{\sigma}_b 
\nonumber\\
 +{ {\rm i}\over{4 m^2}}
(\vec{k}_2 \times \vec{k}_1)_{z} \cdot   (\vec{\sigma}_a)_{z}
+{ {\rm i}\over{4 m^2}}
(\vec{p}_2 \times \vec{p}_1)_{z} \cdot   (\vec{\sigma}_b)_{z} 
\nonumber\\
+{{\rm i}\over{4 m^2}}
(\vec{k}_1)_{z} \cdot \left(\vec{k}_1  \times \vec{\sigma}_a \right)_{z}
-{{\rm i}\over{4 m^2}}
( \vec{k}_2)_{z} \cdot \left( \vec{k}_2 \times \vec{\sigma}_a \right)_{z}
\nonumber\\
+{{\rm i}\over{4 m^2}}
(\vec{p}_1)_{z} \cdot \left(\vec{p}_1 \times \vec{\sigma}_b \right)_{z}
-{{\rm i}\over{4 m^2}}
(\vec{p}_2)_{z} \cdot \left(\vec{p}_2 \times \vec{\sigma}_b \right)_{z}
\ \  \  .
\end{eqnarray}

We can immediately see that the first two terms in $v_{spin}$, which are 
linear in momentum, will lead to difficulties in bound state perturbation 
theory.
In  first-order bound-state perturbation theory they integrate to 
zero, but they enter  at the second-order of bound-state perturbation theory,
 bringing 
the same power of momenta as the familiar term 
$(\vec{q} \times \vec{\sigma}_a)_{\perp}\cdot (\vec{q} \times \vec{\sigma}_b)_{\perp}. $
 So in order to obtain correct splitting of the ground state triplet and 
singlet states using this Hamiltonian, one has to sum  second-order 
bound-state perturbation 
theory using all bound-  and  scattering electron-positron states
 \cite{trulko}.
Let us note that the remaining terms in $(10)$
 would give rise to part of the  spin-orbit 
interactions.

The key to resolving this nuisance is to recognize that the
 spin-independent $v_0$ and the  spin-dependent   $v_{spin}$ are 
both multiplied by the same energy denominators.
We try to find a unitary transformation which, applied to the 
spin-independent term, would generate terms cancelling the unwanted 
linear terms, and restoring rotational invariance in the 
$(\vec{q} \times \vec{\sigma}_a)\cdot (\vec{q} \times \vec{\sigma}_b)$
term.

Consider the following transformation:
\begin{eqnarray}
U_{\alpha} = 1+{{\rm  i}\over{2m}}
\left( \vec{{\cal P}}_{\alpha} \times \vec{\sigma}_{\alpha}\right)_{z}
-{1\over{2}}{ {\cal P}_{\alpha}^{\perp 2}\over{4m^2}}
\end{eqnarray}
for each particle $\alpha$. This transformation is clearly 
unitary to 
second order in momenta, which is all we require here. 
For two particles $a$ and $b$ in the initial and final states,
\begin{eqnarray}
U_{\rm initial}^{\dagger} =
\left[1 - {{\rm  i}\over{2m}}
\left( \vec{k}_{1} \times \vec{\sigma}_{a}\right)_{z} 
-{1\over{2}}{ {k_1}^{\perp 2}\over{4m^2}}
\right]
\left[1- {{\rm  i}\over{2m}}
\left( \vec{p}_{1} \times \vec{\sigma}_{b}\right)_{z}
-{1\over{2}}{ {p_1}^{\perp 2}\over{4m^2}}
\right]  , 
\end{eqnarray}
and
\begin{eqnarray}
U_{\rm final} =
\left[1 + {{\rm  i}\over{2m}}
\left( \vec{k}_{2} \times \vec{\sigma}_{a}\right)_{z} 
-{1\over{2}}{ {k_2}^{\perp 2}\over{4m^2}}\right]
\left[1+ {{\rm  i}\over{2m}}
\left( \vec{p}_{2} \times \vec{\sigma}_{b}\right)_{z}
-{1\over{2}}{ {p_2}^{\perp 2}\over{4m^2}}\right]  . 
\end{eqnarray}

Then the Hamiltonian transforms as:
\begin{eqnarray}
H \rightarrow U_f H U_i^{\dagger}
\end{eqnarray}
leading to new $v_0$ and $v_{spin}$:
\begin{eqnarray}
\tilde{v}_0 = 1 + {1\over{2m^2}}\left(\vec{p}_1 +\vec{p}_2\right)^2
+{1\over{2m^2}}\left(\vec{p_1}^2 +\vec{p_2}^2\right)
\end{eqnarray}
to the leading order, and 
\begin{eqnarray}
\tilde{v}_{spin} = 
& - & {1\over{4 m^2}} 
(\vec{q} \times \vec{\sigma}_a)\cdot 
(\vec{q} \times \vec{\sigma}_b)
\nonumber\\
& + & {3 {\rm i}\over{4 m^2}}
(\vec{k}_2 \times \vec{k}_1) \cdot   \vec{\sigma}_a
+{3 {\rm i}\over{4 m^2}}
(\vec{p}_2 \times \vec{p}_1) \cdot   \vec{\sigma}_b 
\nonumber\\
& - &  {{\rm i}\over{ m^2}}
(\vec{p}_2 \times \vec{p}_1) \cdot  \vec{\sigma}_b
-{{\rm i}\over{ m^2}}
(\vec{k}_2 \times \vec{k}_1) \cdot  \vec{\sigma}_a
\nonumber\\
& + & {{\rm i}\over{4 m^2}}
(\vec{p}_1)_{z} \cdot \left(\vec{p}_1 \times \vec{\sigma}_b \right)_{z}
-{{\rm i}\over{4 m^2}}
(\vec{p}_2)_{z} \cdot \left(\vec{p}_2 \times \vec{\sigma}_b \right)_{z}
 \nonumber\\
& + & {{\rm i}\over{4 m^2}}
(\vec{k}_1)_{z} \cdot \left(\vec{k}_1 \times \vec{\sigma}_a \right)_{z}
-{{\rm i}\over{4 m^2}}
(\vec{k}_2)_{z} \cdot \left(\vec{k}_2 \times \vec{\sigma}_a \right)_{z}
\ \ \   .
\end{eqnarray}
The corrections from energy denominators (i.e. $d$ in eqn. $(8)$) do not 
affect spin-dependent interactions to this order. The term in the unitary 
transformation designed to remove $(\vec{q} \times \sigma)_z$ also removes 
$d (\vec{q} \times \sigma )_z$, because $d$ is an overall factor 
multiplying both $v_0$ and $v_{spin}$.

The rotationally noninvariant terms 
that 
do not mix initial and final state momenta (e.g., 
${{\rm i}\over{4 m^2}}(\vec{p}_1)_{z} \cdot \left(\vec{p}_1 \times \vec{\sigma}_b \right)_{z}$)
 can be removed 
by adding terms of that form into 
the unitary transformation in the second order in momenta:
\begin{eqnarray}
U_{\alpha} \rightarrow U_{\alpha} +
{{\rm i}\over{4 m^2}}({\cal P}_{\alpha})_{z} 
\cdot 
\left( \vec{{\cal P}}_{\alpha} \times \vec{\sigma}_{\alpha}\right)_{z}.
\end{eqnarray}

The resultant spin-dependent interactions is:
\begin{eqnarray}
\tilde{v}_{spin} = 
& - & {1\over{4 m^2}} 
(\vec{q} \times \vec{\sigma}_a)\cdot 
(\vec{q} \times \vec{\sigma}_b)
\nonumber\\
& + & {3 {\rm i}\over{4 m^2}}
(\vec{k}_2 \times \vec{k}_1) \cdot   \vec{\sigma}_a
+{3 {\rm i}\over{4 m^2}}
(\vec{p}_2 \times \vec{p}_1) \cdot   \vec{\sigma}_b 
\nonumber\\
& - &  {{\rm i}\over{ m^2}}
(\vec{p}_2 \times \vec{p}_1) \cdot  \vec{\sigma}_b
-{{\rm i}\over{ m^2}}
(\vec{k}_2 \times \vec{k}_1) \cdot  \vec{\sigma}_a
\end{eqnarray}
which is the familiar Breit-Fermi interaction.

The $\eta_{\mu}\eta_{ \nu}$ term in eqn. $(4)$ which we ignored so far is 
spin-independent and 
already one power of momenta higher than the leading spin-independent term 
in $v_0$. Therefore, to order two powers of momenta higher than the leading 
spin-independent term, it  does not affect 
the spin-spin and tensor interactions. It may affect the spin-orbit 
interactions.
But at least as far as the spin-spin structure, we can now diagonalize the 
new Hamiltonian using states that are related to 
the original states as
\begin{eqnarray}
\vert \tilde{\psi } \rangle = U \vert \psi \rangle \ \ \ .
\end{eqnarray}

It should  be mentioned that the unitary transformation as presented here is 
a next-to-leading order expansion of the  
Melosh transformation \cite{melosh}:
\begin{eqnarray}
{m + x_{\alpha} M_0 
- {\rm i} \left({\cal P}_{\alpha}^{\perp} \times \vec{\sigma}_{\alpha}\right)_{z}
\over{\sqrt{ (m + x_{\alpha} M_0)^2 + {{\cal P}_{\alpha}^{\perp}}^2}}} 
\ \ \  .
\end{eqnarray}

\section{Similarity transformation and fine structure}
In this section we consider corrections that arise due to
 the similarity transformation,
i.e. the $\eta_{\mu} \eta_{\nu}$ term and the corrections due to energy 
denominators in the $g_{\mu \nu}$ term in $(4)$.
For completeness, we mention that the finite cutoffs  also introduce 
corrections, 
 the size of which depends on the specific choice of 
$\Lambda$ \cite{P2}.  

In the previous considerations we omitted corrections due to energy 
denominators in the $g_{\mu \nu}$ term in $(4)$, or $d$ in eqn. $(8)$. 
These corrections do not 
affect the spin-dependent terms, but they do produce a spin-independent 
correction: 
\begin{eqnarray}
4  g^2 (2m)^2\left[ 
{d \over{\vec{q_{ }}^2}} \right] = 
4  g^2 (2m)^2 {1\over{{\vec{q_{ }}}^2 }}
\left[ { \vert q_z \vert \  \vert \vec{q}\cdot
({\vec{p_1}} + {\vec{p_2}} )\vert \over{m \vec{q_{ }}^2}}\right]
\end{eqnarray}
where we have dropped  the omnipresent 
$\sqrt{x_1x_2(1-x_1)(1-x_2)}$ as before. Similarly, any corrections of this 
term due to finite cutoff do not affect spin-dependent interactions.

We now address 
 the $v_{\eta_{\mu} \eta_{\nu}}$ term, and its effect  on 
the spin-orbit interaction. Dropping the omnipresent 
$\sqrt{x_1x_2(1-x_1)(1-x_2)}$, the $\eta_{\mu} \eta_{\nu}$ term gives
\begin{eqnarray}
4 g^2 {1\over{2 (x_1-x_2)^2}}
\left[{
\theta(\vert D_1\vert -\vert D_2\vert )(q^+D_1-q^+D_2)\over{q^+D_1}} 
+ {\theta(\vert D_2\vert -\vert D_1\vert )(q^+D_2 -q^+D_1)\over{q^+D_2}}\right]
 .
\end{eqnarray}

To the lowest order in momenta this equals (for details see appendix):
\begin{eqnarray}
v_{\eta _{\mu} \eta _{\nu}} = 4 g^2 (2m)^2 {1\over{{\vec{q_{ }}}^2 }}
\left[ { \vert q_z \vert \  \vert \vec{q}\cdot
({\vec{p_1}} + {\vec{p_2}} )\vert \over{m q_z^2}}\right] \ \ \ .
\end{eqnarray}

The unitary transformation $(14)$ applied to this term produces \footnote{
This term can affect spin-orbit splittings even if one does not use the 
unitary transformation to rotate the spins. In that case, there would be 
correction in second order bound state perturbation theory, arising from 
the product of $(21)$ with the first two terms in $v_{spin}$ (see 
eqn. $(10)$).}:
\begin{eqnarray}
4 g^2 (2m)^2 {1\over{{\vec{q_{ }}}^2 }}
\left[ { \vert q_z \vert \  \vert \vec{q}\cdot
({\vec{p_1}\over{m}} + {\vec{p_2}\over{m}} )\vert \over{q_z^2}}\right]
\left[1 - {{\rm  i}\over{2m}}
\left( \vec{q} \times \vec{\sigma}_{a}\right)_{z} 
 + {{\rm  i}\over{2m}}
\left( \vec{q} \times \vec{\sigma}_{b}\right)_{z} 
\right] \ \ \ .
\end{eqnarray}

All of these corrections are nonanalytic. This is a consequence of using a 
nonanalytic cutoff function in the similarity transformation.
If the nonanalytic spin-dependent corrections do not vanish, 
a simple the angular momentum operator 
does not emerge even in the nonrelativistic limit.
Fortunately, these spin-dependent terms integrate to zero in the first 
order bound state perturbation theory, since they are odd under parity. 
Terms that appear at higher orders must 
be paired with other terms from second order bound state perturbation 
theory and with terms from higher order similarity transformation.
Corrections which arise due to finite value of cutoff $\Lambda$ do not 
influence the lowest order of spin-dependent interactions for the same reasons.

\section{Conclusions}
We have studied the part of the effective QED Hamiltonian that does not change 
particle number. We have shown that the light-front spin-dependent 
interactions reduce to the familiar Breit-Fermi interactions. This can be 
achieved by a simple unitary transformation corresponding to a change of 
spinor basis. As a consequence of sharp cutoff functions in the similarity 
transformation, there are nonanalytic corrections. These nonanalytic 
corrections produce spin-independent corrections at  ${\cal O}(\alpha ^3)$,
but they do not affect the spin-dependent splittings at order 
${\cal O}(\alpha ^4)$. Photon exchange below the cutoff is needed to remove 
the corrections at order ${\cal O}(\alpha ^3)$.
 
Our primary motivation for restricting the study to the part of Hamiltonian 
which does not change the particle number was QCD. The approach suggested 
by Wilson et al {\cite{thelongpaper} builds on suppressing the exchange of 
low energy gluons by introducing a gluon mass. This makes  higher Fock states 
less important.

As far as aspects discussed here,
in QCD questions about rotational symmetry become more complicated, because 
the finite size of $\Lambda $ comes to play. 
 It 
is straightforward to show that the procedure we outlined here does not 
completely restore manifest rotational invariance in the spin-spin 
interaction in QCD. It is still useful, however, because it helps to separate 
violations of manifest rotational symmetry caused by using light-front spinors
 from the  violations due to  the cutoff \cite{US}. 

\section*{Acknowledgements}
This work
 was supported by the National Science Foundation under grant PHY-9409042. 
M.B. would like to thank R. Furnstahl for 
his valuable comments that helped me to improve this manuscript, and 
S. G{\l}azek for discussions.

\section*{Appendix: $\eta _{\mu} \eta _{\nu} $ term}

Let us concentrate on the expression in the square brackets in eqn. $(20)$.
From eqn. $(6)$ for the energy denominators one can see that the energy 
denominators have form:
\begin{eqnarray}
q^+ D_1& = &a + X \nonumber\\
q^+ D_2& = &a - X \  \  \  ,
\end{eqnarray}
where
\begin{eqnarray}
X=
\left({p_{1  {\rm z}}\over{ m \sqrt{1+{\vec{p_1}^2 \over{m^2}}}}}
-{p_{2  {\rm z}}\over{ m \sqrt{1+{\vec{p_2}^2 \over{m^2}}}}}\right) 
(\vec{p_1}^2 -\vec{p_2}^2) \ \ \ , 
\end{eqnarray}
and 
\begin{eqnarray}
a = -\vec{q_{ }}^2 + {\cal O}(p^6)\ \ \ .
\end{eqnarray}
The difference between the energy denominators is 
\begin{eqnarray}
q^+ D_1 - q^+ D_2 = 2 X \ \ \ .
\end{eqnarray} 
Theta functions can be expressed as
\begin{eqnarray}
\theta(\vert D_1\vert -\vert D_2\vert ) = 
\theta(\vert a+X \vert - \vert a-X \vert) = \theta (a X) \nonumber\\
\theta(\vert D_2\vert -\vert D_1\vert ) = 
\theta(\vert a-X \vert - \vert a+X \vert) = \theta (-a X) \ \ \  .
\end{eqnarray}
Using these expressions, to the lowest nonvanishing order:
\begin{eqnarray} 
\left[ \  {\theta(\vert D_1\vert -\vert D_2\vert )(q^+D_1-q^+D_2)\over{q^+D_1}} 
\right.
& + &\left.
{\theta(\vert D_2\vert -\vert D_1\vert )(q^+D_2 -q^+D_1)\over{q^+D_2}}
\  \right]
\ \ \ \ \ \ \ \ \ \ \ \ \ \ \ 
\nonumber\\
   =   2 \  \left[{
X \theta(a X)\over{a+X}} 
- {X \theta(-a X ) \over{a-X}}\right] 
& \simeq & 2 \  \Big\vert {X\over{a}} \Big\vert 
\  + \  {\cal O}({X^2\over{a^2}}) \ \  
\  . 
\end{eqnarray}
It is easy to show that even if the cutoff $\Lambda $ is kept in place, the 
nonanalytic corrections are still present.

\end{document}